**ORIGINAL ARTICLE**

# Energy-based features and bi-LSTM neural network for EEG-based music and voice classification

Isaac Ariza[1] · Ana M. Barbancho[1] · Lorenzo J. Tardón[1] · Isabel Barbancho[1]

## Abstract
The human brain receives stimuli in multiple ways; among them, audio constitutes an important source of relevant stimuli for the brain regarding communication, amusement, warning, etc. In this context, the aim of this manuscript is to advance in the classification of brain responses to music of diverse genres and to sounds of different nature: speech and music. For this purpose, two different experiments have been designed to acquire EEG signals from subjects listening to songs of different musical genres and sentences in various languages. With this, a novel scheme is proposed to characterize brain signals for their classification; this scheme is based on the construction of a feature matrix built on relations between energy measured at the different EEG channels and the usage of a bi-LSTM neural network. With the data obtained, evaluations regarding EEG-based classification between speech and music, different musical genres, and whether the subject likes the song listened to or not are carried out. The experiments unveil satisfactory performance to the proposed scheme. The results obtained for binary audio type classification attain 98.66% of success. In multi-class classification between 4 musical genres, the accuracy attained is 61.59%, and results for binary classification of musical taste rise to 96.96%.

## 1 Introduction

In daily life, human beings receive information by listening continuously. Information can be due to multiple diverse types of sources: music, speech, natural sounds, etc.; our brain reacts differently to distinct auditory stimuli.

Isaac Ariza, Ana M. Barbancho, Lorenzo J. Tardón, Isabel Barbancho have contributed equally to this work.

✉ Lorenzo J. Tardón
  ltg@uma.es

✉ Isabel Barbancho
  ibp@uma.es

  Isaac Ariza
  iariza@ic.uma.es

  Ana M. Barbancho
  abp@uma.es

[1] ATIC Research Group, ETSI Telecomunicación, Universidad de Málaga, Campus de Teatinos sn, 29071 Málaga, Málaga, Spain

Electroencephalography (EEG) allows getting real-time measurements of how the brain reacts to different stimuli, and, with the help of digital signal processing techniques [1], and machine learning, find out how our brain works. Thus, studies have been carried out to analyze brain responses, and the correlation between audio and EEG data [2]: Darmawan et al. use Electrical Capacitance Volume Tomography (ECVT) to observe the brain's responses to tones of different frequencies with the aim of detecting the area of the brain activated for neuroscience applications [3]; following this line of research, Aggarwal et al. analyze the P300 response of the brain to single-frequency auditory stimuli [4]. Consequently, applications arise from the analysis of EEG signals, like: neurological disorder detection [5, 6], brain computer interfaces (BCI) [7], or emotion and mental task recognition [8–10].

Within this context, different kinds of artificial intelligence techniques have been used for EEG classification, like classic machine learning techniques [11], and neural networks, like convolutional neural networks (CNN) [12, 13] or recursive neural networks (RNN) [14, 15], but other types of neural networks can be used for this kind of analysis, like Cohen–Grossberg neural network [16], and discrete neutral-type neural networks with varying delays [17].

Among previous works in relation to the classification of EEG signals and music, some relevant ones seek to classify music according to the emotions evoked in the subject under test: Seo et al. use the circumplex emotional model to measure responses and examine subjects' emotions based on valence and arousal [18]; different machine learning techniques, support vector machines (SVM), and multilayer perceptron networks (MLP) are used to classify EEG signals into different emotions generated by listening to music with an accuracy of approximately 82% [19]; Poikonen et al. search for brain responses induced by long-lasting pieces of music [20]. Other studies also investigate the effect on the brain of listening to songs in the subjects' native language (Chinese) and a foreign language (Japanese) [21].

Within this context, the aim of this work is to advance in the classification of brain responses to different auditory stimuli using a novel energy-based feature and a bi-LSTM neural network. For this purpose, two different experiments have been specifically designed. In the first experiment, the subject under test listens to songs of different musical genres: ballad, classic, metal, and reggaeton. In the second experiment, sentences in different languages are listened to. The 5 languages selected for this experiment are Spanish, English, German, Italian, and Korean.

EEG signals recorded during these experiments are divided into trials. These trials are processed independently to calculate a measure of energy of each channel, per trial. With these data, matrices of energy differences characterizing each trial are obtained.

With the response of the subjects under test after listening to each song, the brain reaction in terms of energy relations, when listening to a song the subject likes, or dislikes, can be studied. Also, the different behavior when listening to voice or music can also be considered, and even the response to music genres can be observed. To achieve these objectives, binary and multi-class classification tasks with a bidirectional LSTM neural network (bi-LSTM) are carried out, attaining satisfactory performance.

This paper is divided into four parts. First, in Sect. 2, the designed experiments, the dataset employed, and the data acquisition methodology are described. Then, in Sect. 3, the proposed method is exposed, including how EEG recordings are segmented in order to allow the extraction of features; features and the structure of the bidirectional LSTM neural network employed for classification are later described. The results found after the experiments performed, and a discussion on the results is presented in Sect. 4. Finally, some conclusions are drawn in Sect. 5.

## 2 Dataset

This section describes the EEG dataset used in this study; in Sect. 2.1, the EEG acquisition system is presented; next, in Sect. 2.2, the experiments carried out are explained. Finally, in Sect. 2.3, the EEG dataset characteristics are drawn.

### 2.1 EEG recording system

BrainVision actiCHAMP-PLUS system is used to record the EEG signals used in this paper [22]. Sixty-four different channels are controlled with this system in the configuration employed. For each channel, an active electrode, positioned according to the 64-channel arrangement from brain products, is employed. Active electrodes allow high-quality, low-noise recordings. Figure 1 shows the position of the electrodes according to this standard.

The sampling rate is 2500 Hz, and the impedance of the electrodes during recording sessions is below 10 KΩ. Figure 2 includes images of the EEG electrodes, EEG amplifier, recording computer employed, and a subject wearing the electrode helmet. Green light on each electrode indicates that the measured impedance is below 10 KΩ.

Each EEG channel is labeled with a letter and a number. The letter identifies the brain lobe where it is located, and the number indicates the position of the electrode, being odd for the left hemisphere and even for the right hemisphere of the brain. If an electrode is placed on the central line of the head, its number is replaced by letter 'z.'

With this system, channel FCz has been used as reference channel, and channels FT9 and FT10 have been used to capture eye movement. Consequently, these channels are not used in this study, so 61 EEG channels are actually employed.

### 2.2 Experiments

E-prime 3 [23] has been used to design and carry out the experiments. It allows the presentation of multiple auditory and visual stimuli, and collection of feedback from the subject in a synchronous manner.

The experiments were carried out in a room where the test subject was alone, sitting on a chair without armrests, in front of a computer screen, and speakers employed to present stimuli.

Prior to the experiment, the subjects were instructed to relax and avoid movement in order to avoid artifacts in the EEG signals.

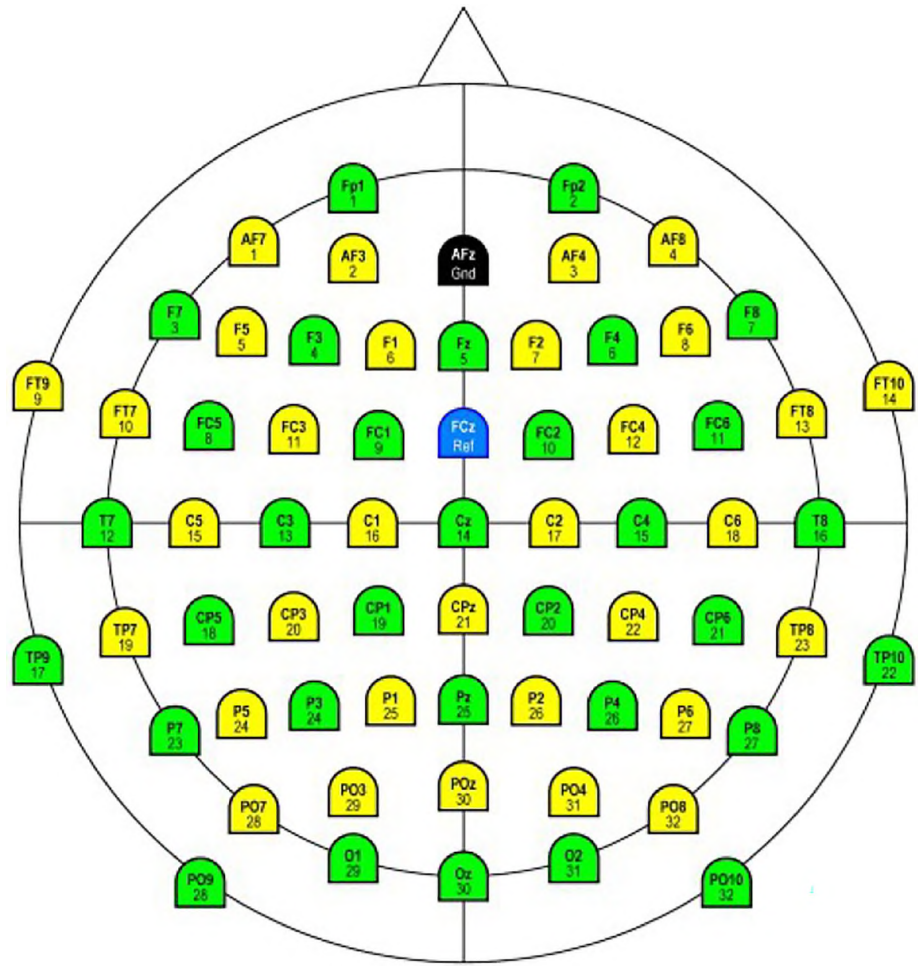

**Fig. 1** Electrode positions according to the 64Ch actiCAP snap standard [22]

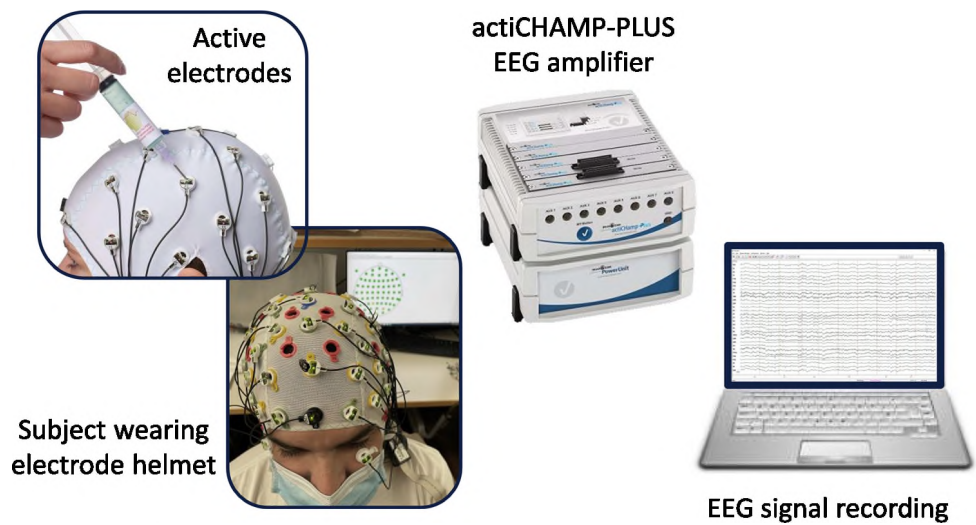

**Fig. 2** Detail of active electrode placement. ActiCHAMP-PLUS EEG signal amplifier, computer used for recording, and a subject under test with the electrode helmet are also shown. Note the green light on the electrodes which indicates that the measured impedance is below 10 KΩ

### 2.2.1 Experiment 1: Listening to fragments of different musical genres

The music experiment consists on listening to fragments of songs from different musical genres. The duration of these fragments is 30 s, and they are extracted from the chorus or the catchiest part of the song. The music genres chosen are: ballad, metal, classic, and reggaeton. These genres have been chosen because they are clearly distinguishable from each other. Five fragments of different songs from each musical genre are played in random order; a total of 20 fragments of 30 s of duration are employed. The global characteristics of the music samples using in this experiment and the labels used to identify the musical genres are shown in Table 1.

The detailed characteristics of the songs used in this experiment are listed in "Appendix."

After listening to each fragment, the test subjects are asked if they knew the song, and if they liked it, with three possible answers: I like it, I like it a bit, or I do not like the song. They are also questioned whether they knew the song beforehand. The possible answers for this question are: I know the song, the song sounds familiar to me, or I do not know the song.

### 2.2.2 Experiment 2: Listening to sentences in different languages

This experiment consists on listening to a series of short sentences, about 5 s of duration, in five different languages: Spanish, English, Italian, German, and Korean. For each language, 30 different sentences are listened by each subject. The sentences are listened to in groups of 5 sentences of each language. The order in which these groups are listened to is random. The global characteristics of the sentences used in this experiment and the labels used to identify the languages are shown in Table 2.

In this study, all the excerpts are simply considered voice samples. The distinction between different languages is beyond the scope of this paper.

## 2.3 Dataset characteristics

The dataset used consists of the recording of EEG signals of 6 healthy people with an average age of 22.3 years, during the realization of the experiments described in Sect. 2.2. Table 3 shows the age and gender of each subject. The native language of all the subjects is Spanish, except for one of them, who is from China, with native language Chinese. Data acquisition methodology was conducted according to the guidelines of the Declaration of Helsinki and approved by Comité de Ética de la Investigación Provincial de Málaga on December 19, 2019, approval number: 2176-N-19.

The duration of experiment 1 (Sect. 2.2.1) is over 9 min, whereas experiment 2 (Sect. 2.2.2) takes 16 min to complete. Signal excerpts are divided into trials of 400ms of duration with an overlap of 50%. Additionally, also trials of 1s of duration with overlap of 50% are considered for evaluation. These trial durations are chosen to capture the P300 generated after the presentation of auditory stimulus. The P300 event is a positive-going potential, peaking at around 300 ms. This component represents the moment when a stimulus causes certain brain response [24]. The two different trial durations (400 ms and 1 s) are considered to observe with which interval the relevant information is better extracted.

For the analysis of results, in the musical genre experiment, the first 30 s of each song was considered, while in experiment 2, the full length of each phrase was used. Table 4 shows the number of each type of trial for experiment 1 and trial length.

**Table 1** Dataset: global characteristics of the music samples used in experiment 1 and labels used to identify musical genres

| Musical genre | Label | # of excerpts | Total duration |
|---|---|---|---|
| Ballad | BA | 5 | 2:30 |
| Classic | CL | 5 | 2:30 |
| Meta | ME | 5 | 2:30 |
| Reggaeton | RE | 5 | 2:30 |

**Table 2** Dataset: global characteristics of the sentences in different languages used in experiment 2 and labels used to identify languages

| Language | Label | # of excerpts | Total duration |
|---|---|---|---|
| Spanish | SP | 30 | 2:20 |
| English | EN | 30 | 2:03 |
| Italian | IT | 30 | 2:30 |
| German | GE | 30 | 1:59 |
| Korean | KO | 30 | 1:46 |

**Table 3** Dataset: subjects participating in the experiments

| Subject # | Age | Gender |
|---|---|---|
| 0009 | 26 | Male |
| 0010 | 20 | Female |
| 0013 | 20 | Female |
| 0014 | 21 | Female |
| 0015 | 22 | Male |
| 0017 | 25 | Female |

**Table 4** Dataset: trials of experiment 1

| Musical genre | Label | # of 400-ms trials | # of 1-s trials |
|---|---|---|---|
| Ballad | BA | 3456 | 1704 |
| Classic | CL | 3456 | 1704 |
| Metal | ME | 3456 | 1704 |
| Reggaeton | RE | 3456 | 1704 |

For the voice/music classification task, samples from experiment 1 are simply considered music excerpts, and samples from experiment 2 are used as voice samples. Table 5 shows the number of trials for each class for both durations: 400 ms and 1 s.

## 3 Analysis methodology

In this section, how data are prepared and processed prior to performing the classification is described. First, the signals used in this study are segmented into 400-ms trials or into 1-s trials as described in Sect. 2.3. Next, we describe the process to obtain the energy-based matrices used to characterize each trial (Sect. 3.1). Finally, in Sect. 3.2, the architecture of the bi-LSTM neural network used to classify the signals is shown.

### 3.1 Data processing

EEG signals obtained using the recording system described in Sect. 2.1 consist of 61 time domain signals, one for each channel. These signals are segmented into 400-ms or 1-s trials with an overlap of 50%. Each of these EEG signals in the time domain will be denoted by $eeg_{n,m}(t)$, which stands for the EEG signal of channel $m$ and trial $n$, with $m$ ranging from 1 to 61.

The representation of each signal in the frequency domain, $EEG_{n,m}(f)$, is obtained by using the fast Fourier transform [25]:

$$EEG_{n,m}(f) = \mathcal{F}(eeg_{n,m}(t)). \tag{1}$$

Then, a measure of the energy per channel and trial is calculated:

$$e_{n,m} = \int_{-\infty}^{\infty} |EEG_{n,m}(f)|^2 \, df, \tag{2}$$

where $e_{n,m}$ represents a measure of the energy of the EEG signal of channel $m$ at trial $n$. Recall that the process can be carried out similarly in the time domain.

Since the amplitude of EEG signals is in the order of microvolts ($\mu V$), then the energy of the signals is low. Consequently, the logarithmic scale is employed for better representation:

**Table 5** Dataset: number of trials of experiment 2

| Class | Label | # of 400-ms trials | # of 1-s trials |
|---|---|---|---|
| Voice | V | 18653 | 8463 |
| Music | M | 13824 | 6816 |

$$E_{n,m}(dB) = 10 \log_{10}(e_{n,m}). \tag{3}$$

Therefore, for each trial $n$, a vector of 61 elements is defined, since 61 EEG signals are considered, one corresponding to each EEG channel:

$$E_n = [E_{n,1}, E_{n,2}, \ldots, E_{n,61}]. \tag{4}$$

Once the measure of energy for each channel, $E_{n,m}$ (dB), is obtained, a matrix that characterizes each trial by calculating the difference, in dB, between the energy in the different channels is obtained. Consequently, a $61 \times 61$ matrix, $\mathbf{E(n)}$, is defined, where the elements of the main diagonal are 0:

$$\mathbf{E(n)} = \begin{pmatrix} E_{n,1} - E_{n,1} & E_{n,1} - E_{n,2} & \cdots & E_{n,1} - E_{n,61} \\ E_{n,2} - E_{n,1} & E_{n,2} - E_{n,2} & \cdots & E_{n,2} - E_{n,61} \\ \vdots & \vdots & \ddots & \vdots \\ E_{n,61} - E_{n,1} & E_{n,61} - E_{n,2} & \cdots & E_{n,61} - E_{n,61} \end{pmatrix}. \tag{5}$$

After $\mathbf{E(n)}$, also the evolution of energy difference between consecutive trials is considered, to observe the temporal evolution of relative energy between the different channels. For this, we use the approximation of the derivative by using the discrete difference [26] for each trial $n$:

$$\mathbf{G'(n)} = \frac{\mathbf{E(n + \Delta)} - \mathbf{E(n)}}{\Delta}. \tag{6}$$

The same expression can be applied to $n - 1$:

$$\mathbf{G'(n-1)} = \frac{\mathbf{E(n-1+\Delta)} - \mathbf{E(n-1)}}{\Delta}. \tag{7}$$

Then, it is possible to define a centered approximation of the derivative of the energy of trial $n$ by using the average of the previous expressions. Also, let $\Delta = 1$, then:

$$\mathbf{E'(n)} = \text{mean}(\mathbf{G'(n)}, \mathbf{G'(n-1)})_{\Delta=1} = \frac{\mathbf{G'(n)} + \mathbf{G'(n-1)}}{2}, \tag{8}$$

where $\mathbf{E'(n)}$ represents a matrix defined upon the selected approximation of the derivative of the energy of trial $n$.

### 3.2 Bidirectional LSTM neural network

Among the different types of neural networks most commonly used, a long short-term memory (LSTM) neural network [27, 28] has been selected. This kind of neural network makes use of temporal dependencies in the signal, creating context for its performance [29]; this is accomplished by forgetting irrelevant information, while storing relevant parts of novel information to update its internal state. LSTM networks and its derivatives may be the most commonly used type networks employed in EEG-based emotion recognition [29]. Specifically, in our case, a bidirectional LSTM neural network, bi-LSTM, has been chosen, since the identification between voice and music, as well as the identification of musical genres, requires both previous and posterior context [30], and musical taste can, on the other hand, be considered connected to emotions.

The neural network used to perform the classification tasks consists of a set of layers that receive an input, process the data, and generate an output compatible with the next layer. First, a sequence input layer is used to introduce the characterization of trials into the neural network. The network has 61 inputs, one for each EEG channel. Next, a bidirectional LSTM layer is used to process the input data and select the relevant information. Then a fully connected layer is used that additionally defines the number of outputs the neural network will have. Finally, a nonlinear softmax layer and a classification layer are employed. The whole processing scheme is illustrated in Fig. 3.

The number of inputs of the neural network, $N$, is the number of EEG channels (61). Each input is a row of matrix $\mathbf{E}(\mathbf{n})$ or $\mathbf{E}'(\mathbf{n})$, which are described in Sect. 3.1. The number of outputs depends on the type of test carried out. For binary classification task, the number of outputs is 2, and for multi-class classification tasks it is the number of classes of each task: 3 and 4 in our case.

## 4 Results

In this section, the results obtained for the different experiments carried out are shown. Evaluations have been performed for inter-subject scenario using the data available. Data collected from experiment 1 (Sect. 2.2.1) and experiment 2 (Sect. 2.2.2) allow performing different tests. First, using the data of both experiments, a binary classification aimed at the discrimination between two classes: voice (V) and music (M) have been carried out. In Sect. 4.1, $F$-measures attained are shown.

Then, using the data of experiment 1, a multi-class classification task is defined regarding musical genres. The results are shown in Sect. 4.2.

Finally, Sect. 4.3 displays the results of binary and multi-class classification regarding musical taste.

Note that for all the tests, the neural network has the same configuration: a bi-LSTM layer is configured with 20 hidden units, and the network is trained for 5 epochs. The dataset used in each test is divided randomly into two subsets, one subset with 50% of the trials to train the network, and the other 50% of the trials are used to test the performance.

Confusion charts [31] have been used to visualize the results obtained and to evaluate the performance [32] of the classification system proposed.

### 4.1 Binary classification: spoken voice and music

In these tests, the objective is to distinguish between two different situations: when listening to music and when listening to speech. To this end, data gathered from both experiments, 1 and 2, have been used. First, trials of experiment 1, that correspond to the data collected when the subjects are listening to music, are labeled as music. 'M' is used as the label of this class. Then, in experiment 2, subjects are listening sentences in different languages; these data are labeled as spoken voice; 'V' is used as label for these samples.

In order to perform the binary classification tests, the dataset is split into two even subsets: the train subset, used to train the neural network, and the test subset, used to evaluate the performance. Note that unbalanced classes are often a drawback for classification and evaluation, so we use under sampling to use the same number of music and voice samples: 16018 trials of each.

Every test is carried out 10 times, randomly changing the trials used to train and test the network. The results shown are the average of these 10 evaluations. Figure 4 shows the confusion chart using trials of 400 ms of duration with $\mathbf{E(n)}$ used to characterize each trial. The results for different scenarios evaluated are exposed in Table 6.

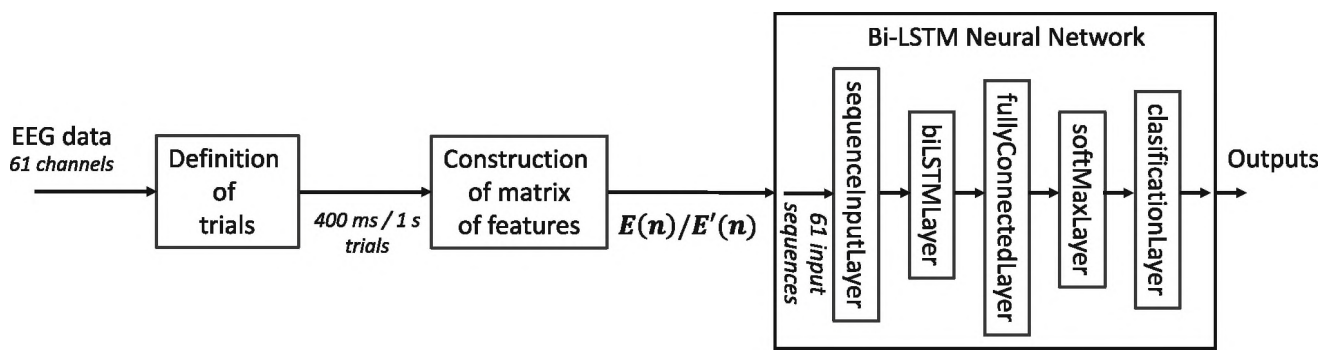

Fig. 3 Overview of the whole EEG signal processing scheme

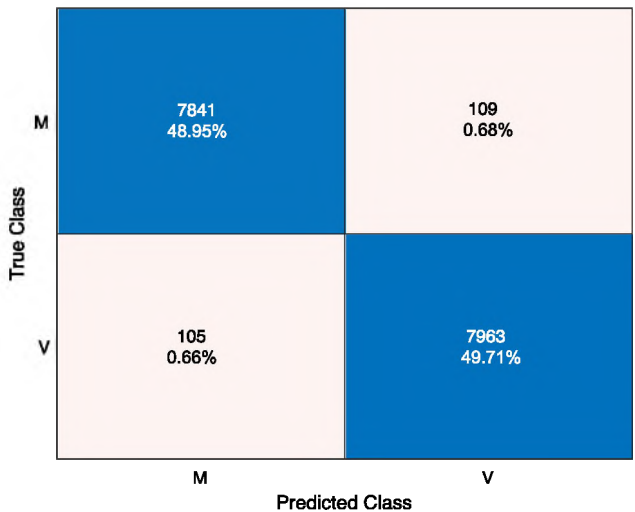

**Fig. 4** Confusion chart for binary classification using EEG data of subjects listening to music (M) or spoken voice (V). $\mathbf{E(n)}$ was used to characterize each trial of 400 ms of duration

### 4.2 Multi-class classification: musical genres

Apart from the binary classification tests to distinguish between spoken voice and music, a multi-class classification task has been considered on the basis of the same characterization scheme and classification method.

Data gathered from experiment 1 (Sect. 2.2.1) are used to carry out a 4-class classification task, where a class is defined according to the musical genre.

Two different tests have been carried out in this multi-classification task scenario: first, the classification is performed using the matrix $\mathbf{E(n)}$; then, with the aim of further considering the temporal evolution of the energy in different channels, $\mathbf{E'(n)}$ is used to characterize each trial.

The dataset was split into two even subsets with elements randomly chosen, as explained in Sect. 4.1. Each test was carried out 10 times; the results shown are the average of all the evaluations performed.

The accuracy attained in the first test, using $\mathbf{E(n)}$ to characterize the EEG trials within the bi-LSTM network, is 53.40%.

In the second test, using $\mathbf{E'(n)}$, the accuracy obtained is 61.59%. Figure 5 shows the confusion chart corresponding to this test.

### 4.3 Test to evaluate whether a subject likes a song, or not

Using the data obtained after experiment 1, and the subjects' responses, the trials are labeled depending on their musical taste: if the subject likes the song, it is labeled as 'L,' if they like it a bit, it is labeled as 'B' and if they did not like the song, it is labeled as 'NL.' With this, two different tests have been carried out:

1. Binary classification: like/dislike.
2. Multi-class classification: like/like a bit/dislike.

These tests have been performed with trials of 400 ms with an overlap of 50% segmented as described in Sect. 3.1 and also with trials of 1 s with an overlap of 50%. Also, both characterization matrices, $\mathbf{E(n)}$ and $\mathbf{E'(n)}$ are employed. The results obtained are shown in the following subsections.

#### 4.3.1 Binary classification: like or dislike

In this test, the brain reaction depending on the subject's musical taste when listening songs from different musical genres is observed. This is done by performing a binary classification task with two different classes: the subject likes the song ('L') and the subject does not like the song ('NL'). The results for the different scenarios evaluated are exposed in Table 7.

The highest accuracy is attained when the classification task is performed using the matrix $\mathbf{E(n)}$ for trials of 400 ms. Figure 6 shows the confusion chart for this scenario.

#### 4.3.2 Multi-class classification: like/like a bit/dislike

In this experiment, a multi-class classification task is performed according to subject's musical taste with three possible classes: the subject likes the song, labeled as 'L,'

**Table 6** Results of binary classification: music or spoken voice

| Matrix | Duration | Accuracy (%) | F-score (%) | Recall (%) | Precision (%) |
|---|---|---|---|---|---|
| $\mathbf{E(n)}$ | 400 ms | 98.66 | 98.65 | 97.90 | 98.63 |
| $\mathbf{E(n)}$ | 1 s | 98.81 | 98.82 | 99.62 | 98.04 |
| $\mathbf{E'(n)}$ | 400 ms | 99.90 | 99.90 | 99.81 | 100 |
| $\mathbf{E'(n)}$ | 1 s | 99.97 | 99.97 | 99.94 | 100 |

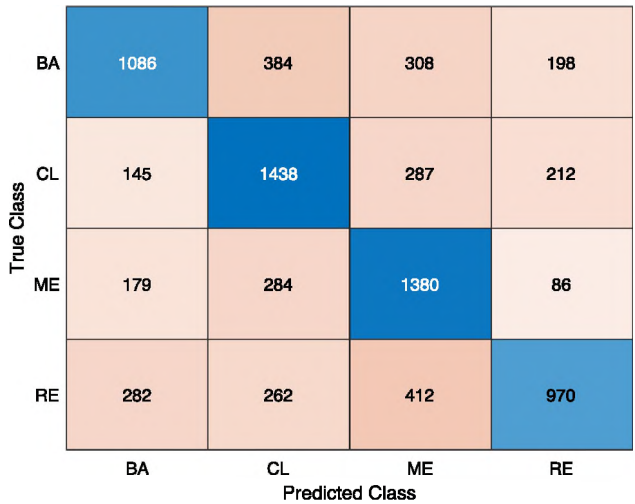

Fig. 5 Confusion chart for multi-class classification using EEG data of subjects listening to music of different musical genres. $\mathbf{E'(n)}$ is employed

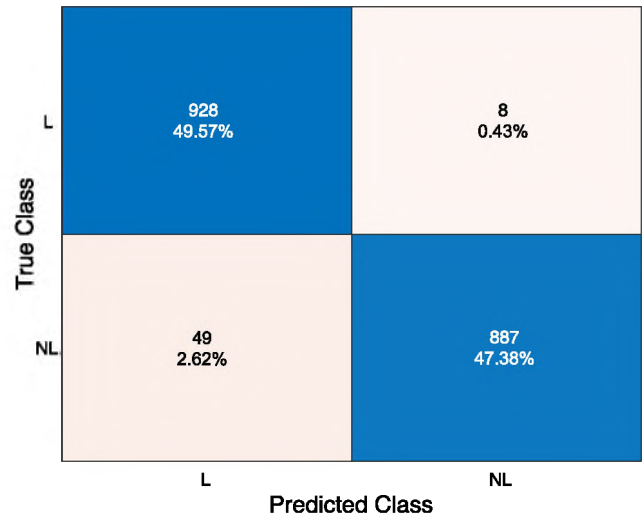

Fig. 6 Confusion chart for binary classification regarding musical taste using EEG data of subjects listening to songs of different musical genres. In this experiment, $\mathbf{E(n)}$ obtained for trials of 400 ms is used

the subject likes the song a bit, labeled as 'B,' and the subject does not like the song, labeled as 'NL.' The accuracy attained in the different scenarios evaluated is exposed in Table 8.

The highest accuracy is attained when the classification task is performed using the matrix $\mathbf{E(n)}$ for trials of 400 ms. Figure 7 shows the confusion chart for this scenario.

### 4.4 Discussion

The aim of this work is to observe how the human brain responds differently to different auditory stimuli, and use such behavior to define characterization, and classification schemes for identification tasks regarding different aspects of audio listening on the basis of EEG signals, and their automatic analysis.

In pursuit of this goal, a characterization scheme based on energy relations between EEG signals acquired at different locations of the brain, and a classification scheme working on two types of tests: binary and multi-class classification, have been defined.

In the first tests, two different but related types of stimuli were considered: spoken voice and music. At the sight of the results obtained, Table 6, it can be considered that there is evidence that the brain responds differently to these stimuli, and the differences are captured by measurements of energy relations at different locations of the brain. Specifically, the relations between the derivative of energy yield the best results; note that the choice of this feature adds an additional specific layer to consider temporal evolution of energy.

Also, a trend to attain the best results with longer trials can be discovered, which means differences are slightly more easily identified in such case. Note that results are similar for both $\mathbf{E(n)}$ and $\mathbf{E'(n)}$ which can be due to the fact that the network selected makes use of temporal context to perform. Figure 8 shows the topoplot of Subject 0017 while listening to speech (voice) and while listening to music. In this figure differences regarding energy distribution of EEG signals across the brain are observed.

Table 8 Results of multi-class classification: like/like a bit/dislike

| Matrix | Duration | Accuracy (%) |
|---|---|---|
| $\mathbf{E(n)}$ | 400 ms | 92.41 |
| $\mathbf{E(n)}$ | 1 s | 91.20 |
| $\mathbf{E'(n)}$ | 400 ms | 92.24 |
| $\mathbf{E'(n)}$ | 1 s | 91.02 |

Table 7 Results of binary classification: the subject likes or dislikes a song

| Matrix | Duration | Accuracy (%) | F-score (%) | Recall (%) | Precision (%) |
|---|---|---|---|---|---|
| $\mathbf{E(n)}$ | 400 ms | 96.96 | 97.02 | 99.15 | 94.98 |
| $\mathbf{E(n)}$ | 1 s | 95.89 | 95.93 | 96.97 | 94.92 |
| $\mathbf{E'(n)}$ | 400 ms | 96.32 | 96.38 | 98.16 | 94.67 |
| $\mathbf{E'(n)}$ | 1 s | 95.43 | 95.44 | 95.55 | 95.33 |

Fig. 7 Confusion chart for multi-class classification regarding musical taste using EEG data of subjects listening to songs of different musical genres. In this experiment, **E(n)** obtained for trials of 400 ms is used

With respect to the multi-class case, tests have been carried out to evaluate, again, the utilization of energy measures within the bi-LSTM network to analyze brain responses when listening to songs of different musical genres, and, also, regarding the subjects' musical taste. It must be noted that the results obtained regarding musical genre improve when specific temporal energy evolution is introduced in the study by considering a measure of difference in energy relations between consecutive samples. In these tests, the results obtained classifying musical genres attained an accuracy of 61.59%.

Figure 9 represents the energy of acquired EEG signals in the different areas of the brain, over a diagram of the head. There are 4 different graphs, each one representing the average of the 4 trials of each gender for subject 0017. We can see that activation areas mostly coincide; however, levels change depending on the genre of music being listened to.

Regarding musical taste, according to results shown in Sect. 4.3, the brain reacts in different ways depending on whether the subject likes the song, or not, or they like it a bit, and this fact can be observed through the analysis of energy relations across diverse brain areas. The accuracy attained is higher than 90% in all the scenarios evaluated and over 95% in the case of the binary decision task.

The best results for these experiments, both binary and multi-class classification tasks, are achieved when **E(n)** obtained for 400-ms EEG trials is used. Nevertheless, the performance attained in different configurations regarding energy is similar; this can be due to the fact that temporal evolution is already captured by the neural network specifically chosen for its capability to make use of temporal context in its performance.

## 5 Conclusion

In this paper, an analysis of brain responses to different auditory stimuli we receive on a daily basis has been carried out. For this purpose, two experiments were designed in which subjects listen to songs of diverse genres and phrases in different languages. Then, in order to characterize the brain response in terms of EEG signals, a matrix with relative energy measures between the different channels was defined. Also, a bi-LSTM neural network was used to classify the different samples in several classification tasks defined. Time evolution of the energy of EEG signals was taken into account by mean of the neural network architecture selected, and, also, by using an approximation of the derivative of the energy measures considered.

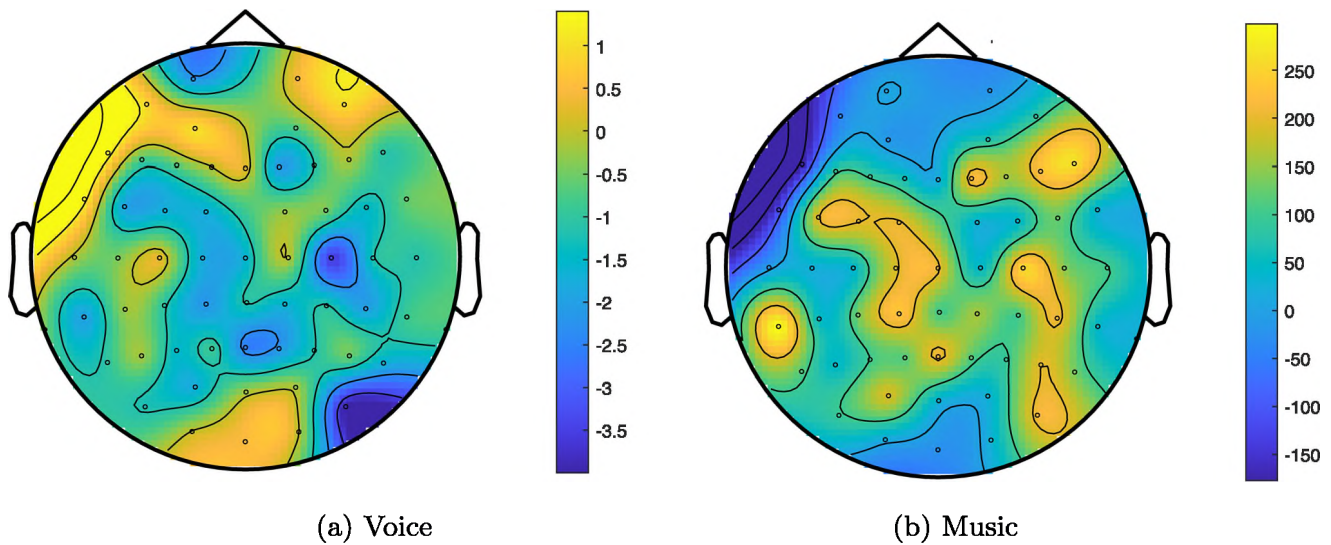

(a) Voice  (b) Music

Fig. 8 Topoplot of Subject 0017 while listening to speech (voice) and while listening to music

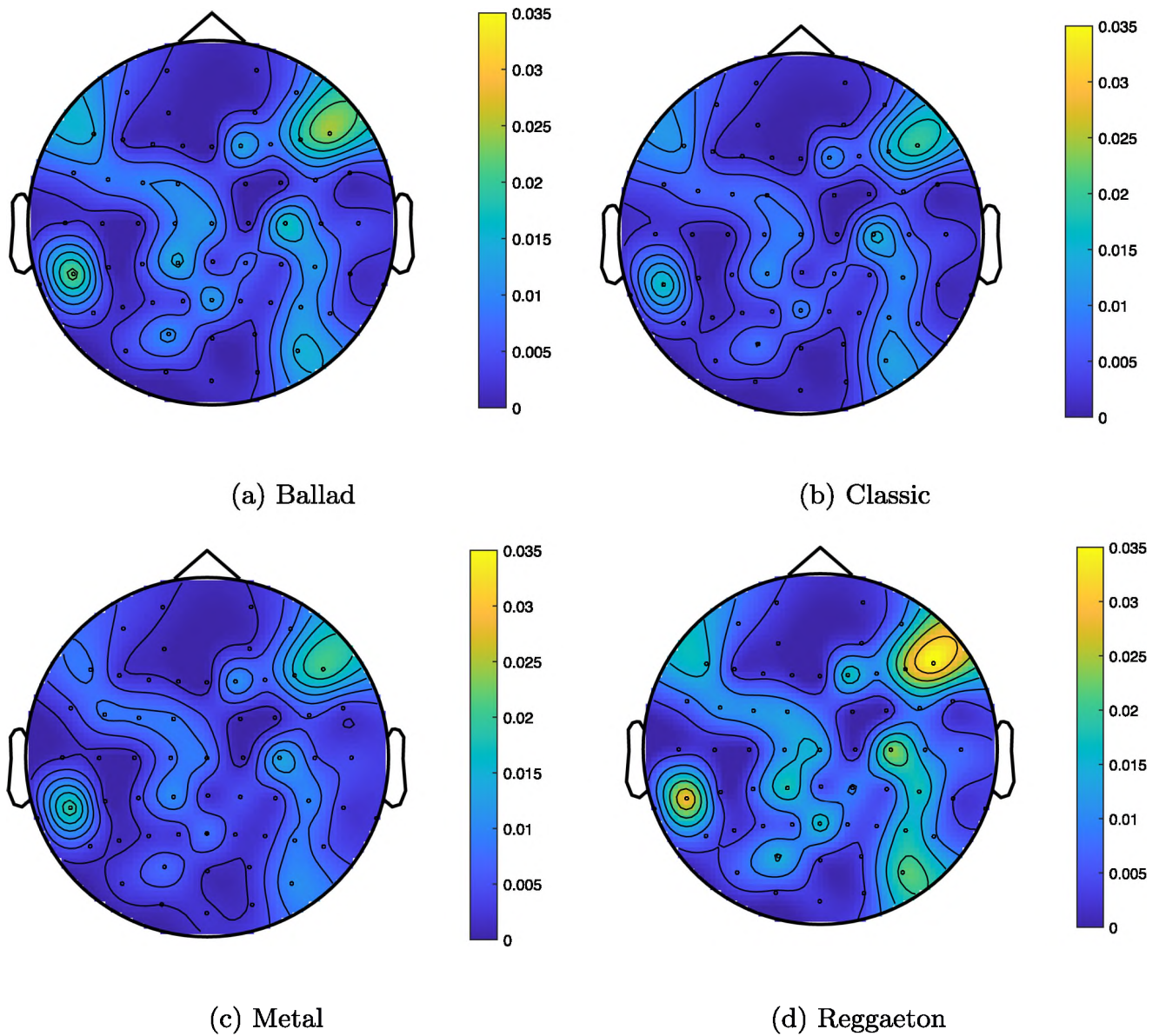

Fig. 9 Average of the energy of 4 trials of each musical genre for Subject 0017

The scheme and features implemented were used to perform two kinds of classification tasks. On one hand, binary classification to distinguish between trials defined while the subjects are either listening to music or listening to speech was carried out. High accuracy was attained in this task with the energy-based features and classification scheme selected: over 98% of accuracy in all cases.

On the other hand, multi-class classification task was carried out for the experiment consisting on listening to songs of varied genres. EEG signals were classified into 4 different groups, one for each musical genre. Then, the classification stage was performed using the energy-based measures defined for the acquired EEG signals within the classification structure selected. The accuracy attained was 61.59%, this lower performance with respect to the previous cases can be due to the fact that the task, in all cases now, is actually the same; nevertheless, it is still possible to perform the classification task to some extent using measures of energy.

Also, classification experiments were carried out regarding musical taste. In this case, the accuracy attained was high, over 90%, in both the binary and multi-class schemes. This can be due to the distinct brain reaction in the different situations regarding taste. This sound result could be specifically useful for the development of music recommendation applications.

The next steps in this study could be varied regarding diverse types of sounds and tasks; however, also, the analysis of reduction of the number of EEG channels to perform this kind of task could help to make the approach more widely available.

# Appendix: Songs used in Experiment 1

See Tables 9, 10, 11, and 12.

Table 9 Dataset Experiment 1: Ballad

| Song | Singer | Duration (s) |
| --- | --- | --- |
| Una estrella en mi jardín | Mari Trini | 30 |
| Libre | Nino Bravo | 30 |
| I will always love you | Whitney Houston | 30 |
| Yesterday | The Beatles | 30 |

Table 10 Dataset Experiment 1: Classic

| Name of file | Database | Duration (s) |
| --- | --- | --- |
| classical.00011 | GTZAN [33] | 30 |
| classical.00089 | GTZAN [33] | 30 |
| classical.00094 | GTZAN [33] | 30 |
| classical.00096 | GTZAN [33] | 30 |

Table 11 Dataset Experiment 1: Metal

| Name of file | Database | Duration (s) |
| --- | --- | --- |
| metal.00007 | GTZAN [33] | 30 |
| metal.00046 | GTZAN [33] | 30 |
| metal.00070 | GTZAN [33] | 30 |
| metal.00092 | GTZAN [33] | 30 |

Table 12 Dataset Experiment 1: Reggaeton

| Song | Singer | Duration (s) |
| --- | --- | --- |
| Despacito | Luis Fonsi | 30 |
| Gasolina | Daddy Yankee | 30 |
| La gozadera | Gente de Zona, Marc Anthony | 30 |
| Danza Kuduro | Don Omar | 30 |

**Author Contributions** All authors contributed to the study conception and design. Material preparation, data collection, analysis, and writing were performed by IA, AMB, LJT, and IB. All authors read and approved the final manuscript.

**Funding** Funding for open access publishing: Universidad Málaga/CBUA. This publication is part of Project PID2021-123207NB-I00, funded by MCIN/AEI/10.13039/501100011033/FEDER, UE. This work was partially funded by Junta de Andalucía, Proyectos de I+D+i, in the framework of Plan Andaluz de Investigación, Desarrollo e Innovación (PAIDI 2020), under Project No. PY20_00237. Funding for open access charge: Universidad de Málaga/CBUA. This work was done at Universidad de Málaga, Campus de Excelencia Internacional Andalucia Tech.

# Declarations

**Conflict of interest** The authors declare no conflict or competing interests.

**Ethics approval** Data acquisition methodology was conducted according to the guidelines of Declaration of Helsinki and approved by Comité de Ética de la Investigación Provincial de Málaga on December 19, 2019, approval number: 2176-N-19.

**Consent to participate** All the participating subjects gave their written consent.

**Availability of data and materials** Data will be made available upon reasonable request and nondisclosure agreement.

**Code availability** Code will be made available upon reasonable request.